# Neuro-imagerie néonatale : quelle valeur prédictive ?
## Jessica Dubois, Lucie Hertz-Pannier, Sara Neumane

**Affiliations**

1. Université Paris Cité, Inserm, NeuroDiderot, F-75019 Paris, France
2. Université Paris-Saclay, CEA, NeuroSpin, UNIACT, F-91191, Gif-sur-Yvette, France

La naissance prématurée ainsi que différentes agressions pré- et périnatales peuvent entraîner des lésions cérébrales diverses et ont clairement été identifiées comme des facteurs de risque majeurs de troubles neurodéveloppementaux, avec des conséquences variables mais multiples, qui peuvent altérer significativement le devenir fonctionnel des enfants à court, moyen et long terme. Les différents marqueurs diagnostiques et pronostiques disponibles, fondés notamment sur des données cliniques et de neuro-imagerie, ont comme principal objectif de faciliter le dépistage précoce des différents troubles et d'optimiser le suivi et la prise en charge de ces bébés à risque de déficiences.

Les dernières décennies, marquées par d'importantes avancées des techniques de neuro-imagerie dont l'imagerie par résonance magnétique (IRM), ont permis de développer des protocoles d'acquisition dédiés à cette population ainsi que des méthodes d'analyse spécifiques au cerveau en développement. Différentes études ont ainsi confirmé l'apport indiscutable de l'IRM pour la caractérisation fine des lésions cérébrales en période périnatale, et ont exploré le potentiel diagnostique et pronostique de cette technique. De plus, le caractère non invasif et non irradiant de l'IRM permet un suivi longitudinal et la mise en évidence des trajectoires développementales.

## Explorer les relations entre la structure et la fonction cérébrales par IRM

Les différentes séquences d'IRM fournissent des marqueurs reflétant les caractéristiques anatomiques et physiologiques du système nerveux central, et d'importantes relations structure-fonction ont récemment été établies dans le cerveau en développement.

Au-delà de la visualisation des lésions et des anomalies de signal au sein des structures cérébrales (liées à une gliose cicatricielle, un retard de myélinisation, etc.), l'IRM anatomique pondérée en T1 et T2, couplée à des post-traitements d'images spécifiques (comme la segmentation tissulaire), permet de mesurer les volumes (globaux ou régionaux) de substance grise corticale et sous-corticale, de substance blanche et de liquide cérébrospinal (LCS). Il est également possible d'analyser quantitativement la forme des structures cérébrales et le plissement cortical (sillons, gyri) avec des approches de morphométrie.

Utilisée classiquement pour détecter les lésions ischémiques aiguës, l'IRM de diffusion est sensible à la diffusion des molécules d'eau et fournit des mesures d'intégrité tissulaire et de microstructure via des paramètres quantitatifs, comme le coefficient de diffusion moyen et l'anisotropie fractionnelle obtenus en imagerie du tenseur de diffusion (DTI). Ces paramètres varient avec la myélinisation de la substance blanche et sont corrélés avec la maturation fonctionnelle des réseaux, comme cela a été montré avec le lien entre les capacités visuelles des bébés (nés prématurément ou non) et le stade de maturation des voies optiques par exemple. Grâce aux approches de tractographie, l'IRM de diffusion permet aussi de mettre en évidence les principaux faisceaux de substance blanche et d'analyser la connectivité structurelle en développement.

D'autres techniques IRM commencent à être disponibles en clinique pour explorer le cerveau du bébé. La relaxométrie IRM mesure les temps de relaxation T1 et T2 qui reflètent la maturation cérébrale et la myélinisation. L'IRM pondérée en susceptibilité magnétique (SWI) permet de détecter la présence de lésions hémorragiques même minimes. Certaines techniques analysent la perfusion cérébrale de façon non invasive avec un marquage magnétique (*labelling*) des spins artériels (ASL).

La spectroscopie par résonance magnétique (SRM) permet d'étudier le métabolisme cérébral en mesurant des métabolites tels que le N-acétyl-aspartate (NAA, impliqué dans la viabilité neuronale), la créatine (Cr, nécessaire à la régulation énergétique des cellules), la choline (Cho, marqueur du renouvellement des membranes), le myo-inositol (mI, marqueur glial), le complexe glutamate-glutamine (Glx, impliqué dans les processus de maturation et de destruction des neurones) et le lactate (Lac, marqueur d'hypoxie ou de déficit énergétique des cellules).

Enfin, les techniques d'IRM fonctionnelle (d'activation ou au repos), actuellement disponibles pour cet âge en recherche uniquement, peuvent informer sur le développement fonctionnel du cerveau et sur la connectivité fonctionnelle entre réseaux neuronaux.

Ces multiples techniques IRM permettent donc une caractérisation non invasive et relativement fine du développement cérébral ainsi que d'anomalies variées chez le nouveau-né à risque de troubles du neurodéveloppement.

**Neuroplasticité et lésions cérébrales en période périnatale**

Dans les suites d'une lésion, même précoce, le système nerveux central ne peut se réparer *ad integrum*, mais il existe des mécanismes permettant des modifications et adaptations plus ou moins fonctionnelles des réseaux concernés. Cette plasticité post-lésionnelle est par nature dépendante des caractéristiques lésionnelles, mais aussi du degré de maturation — au moment de la lésion — des réseaux cérébraux endommagés, ainsi que des stimulations qu'ils pourront recevoir par la suite. Le potentiel de récupération fonctionnelle est surtout attribué à la mise en jeu précoce des circuits fonctionnels non endommagés et en cours de développement.

Par ailleurs, le principe de *bonne récupération précoce* — principe de Kennard : concept selon lequel le cerveau immature serait plus résistant aux agressions que le cerveau mature, grâce à une plasticité qui lui permettrait de se remodeler et se réorganiser plus efficacement pour préserver les fonctions — est aujourd'hui nuancé par la notion de *vulnérabilité précoce*. En effet, le cerveau immature présenterait une vulnérabilité intrinsèquement plus grande à certaines agressions (exposition prénatale à l'alcool, hypoxie néonatale et dommages de l'hippocampe, etc.). De plus, les différents réseaux cérébraux maturent à des rythmes différents et selon une séquence particulière avec, très schématiquement, en premier les réseaux primaires (sensori-moteurs, visuels, auditifs… pendant la grossesse et les premiers mois postnataux), suivis de près par les régions associatives unimodales, puis par les régions associatives multimodales (comme les régions pariétales postérieures et préfrontales) dont la maturation s'étend jusqu'à la fin de l'adolescence. Suite à une agression cérébrale précoce, ce sont les réseaux les plus matures qui montrent la plus faible plasticité. Une lésion impactant un réseau donné vient perturber non seulement la fonction normalement sous-tendue par ce réseau (en altérant des circuits déjà opérationnels, même si encore immatures), mais aussi la séquence neurodéveloppementale globale subséquente du fait de l'altération du développement d'autres régions, adjacentes ou associées d'un point de vue fonctionnel, et de l'interdépendance des différentes fonctions cérébrales au cours du développement.

Ainsi, suite à une atteinte précoce des réseaux primaires, les déficiences pourront concerner, dans les premiers mois, principalement les acquisitions sensorielles et motrices (posturales et fines). Puis progressivement des troubles (apparition retardée, incomplète, voire absence des acquisitions attendues pour l'âge de développement de l'enfant) pourront se manifester dans d'autres domaines comme les interactions sociales, la communication, l'adaptation émotionnelle et comportementale. D'éventuelles difficultés cognitives et troubles des apprentissages pourront se révéler à l'âge scolaire.

## Quelle valeur pronostique de l'IRM précoce ?

Dans une démarche thérapeutique et de suivi, les cliniciens s'interrogent sur la valeur pronostique de l'IRM réalisée précocement chez le nouveau-né suite à une agression pré- ou périnatale, puisque le risque de troubles neurodéveloppementaux n'est qu'en partie dépendant des caractéristiques de la lésion cérébrale.

De façon schématique, on peut distinguer trois contextes cliniques (Figure 1) :

- lorsque l'imagerie réalisée à l'âge équivalent du terme de la grossesse est *normale*, de façon concordante avec l'examen clinique et en l'absence de facteurs de risque majeurs, le pronostic est bon : l'IRM montre alors une bonne valeur prédictive négative ;
- à l'opposé, le pronostic est très péjoratif dans certaines conditions cliniques « *catastrophiques* » où les données de neuro-imagerie montrent clairement une atteinte cérébrale majeure en lien avec une symptomatologie précoce spécifique, pouvant parfois amener à une décision de limitation ou d'arrêt des thérapeutiques actives ;
- dans la majorité des *cas intermédiaires*, la situation est plus complexe, avec des anomalies cérébrales plus subtiles, non encore détectables (en tout cas avec les techniques utilisées en routine clinique) ou responsables de signes cliniques d'apparition décalée dans le temps. Il est alors nécessaire de prendre en compte les différents éléments mis à disposition (anamnèse, suivi clinique et neurologique, neuro-imagerie précoce) pour converger vers un pronostic réaliste et organiser la prise en charge spécifique du nouveau-né. Les différentes techniques d'IRM sont alors particulièrement pertinentes puisqu'elles fournissent des marqueurs variés du développement anatomofonctionnel du cerveau. De façon schématique, la valeur pronostique de l'IRM sera meilleure dans les atteintes fonctionnelles à développement précoce (systèmes primaires), que dans les atteintes à développement plus tardif (systèmes associatifs, fonctions cognitives élaborées).

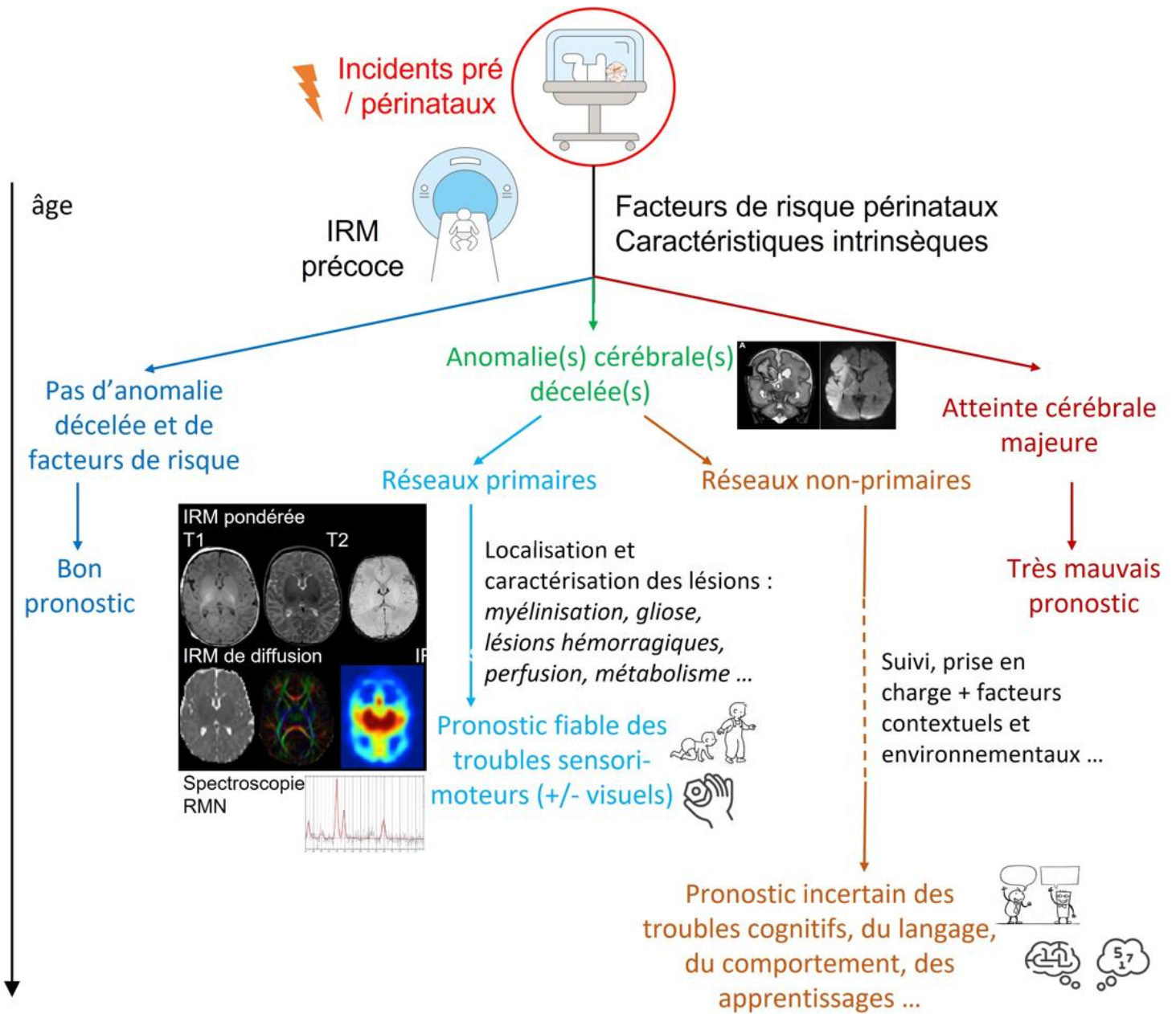

*Figure 1* : *Valeur pronostique de l'IRM précoce suite à une agression pré- ou périnatale, en fonction du contexte clinique, des réseaux cérébraux atteints et de l'âge de l'enfant*

**Pronostic moteur et visuel**

Lorsque la lésion périnatale touche significativement les réseaux sensorimoteurs primaires, une forte corrélation est observée entre les caractéristiques de la lésion et l'évolution fonctionnelle, avec des déficits moteurs et/ou sensoriels pouvant aller jusqu'à la paralysie cérébrale.

Dans le contexte d'un **accident vasculaire cérébral (AVC) néonatal**, la topographie (localisation et étendue) de la lésion permet ainsi de prédire le devenir moteur. Un AVC sylvien complet entraîne quasi systématiquement une hémiplégie controlatérale (de l'hémicorps droit pour un AVC gauche, et réciproquement). Pour les AVC plus localisés, il convient d'évaluer en IRM l'atteinte des faisceaux cortico-spinaux moteurs ipsilatéral et controlatéral à la lésion, ainsi que les atteintes des noyaux gris centraux et des thalamus, pour mieux prédire les déficiences motrices. Des anomalies de signal pondéré T1, T2 et/ou de microstructure en IRM de diffusion au niveau du bras postérieur de la capsule interne et/ou du pédoncule cérébral font suspecter une évolution vers une paralysie cérébrale unilatérale spastique (avec même une valeur prédictive positive de 100 % en cas d'atrophie du pédoncule).

Dans le contexte d'une **encéphalopathie hypoxique ischémique chez le nouveau-né à terme**, une mauvaise évolution motrice est également prévisible en cas d'anomalies IRM (dont une restriction de la diffusion) au niveau des noyaux gris centraux, des thalamus ou des capsules internes.

En cas de **prématurité**, les conditions les plus à risque de déficiences motrices sont les leucomalacies périventriculaires (LPV) kystiques de grade III (marche autonome dans seulement environ 10 % des cas), les hémorragies intraventriculaires (HIV) de grade III ou IV (risque de paralysie cérébrale : environ 25 % et 50 % respectivement), et les lésions hémorragiques significatives du cervelet. Les marqueurs pronostiques obtenus en IRM à l'âge équivalent du terme, portent sur des anomalies de la substance blanche impliquant le faisceau corticospinal (bras postérieur de la capsule interne) et les radiations thalamiques (au niveau du trigone), ainsi que des noyaux gris centraux et des thalamus. D'autres anomalies de la substance blanche seraient aussi associées à des déficiences motrices mais de façon moins systématique : les lésions punctiformes de la substance blanche périventriculaire (de nature ischémique ou hémorragique) et les anomalies DEHSI (intensité du signal pondéré en T2 excessivement élevée et diffuse dans la substance blanche).

Dans ces trois conditions périnatales (AVC, encéphalopathie, prématurité), il est également possible de pronostiquer, dans une certaine mesure, des anomalies du devenir visuel lorsque des anomalies en IRM anatomique et de diffusion sont observées au niveau des réseaux visuels primaires (radiations optiques, cortex occipital…). Néanmoins une déficience visuelle sévère ne semble pas systématiquement résulter d'une atteinte de ces régions.

## Pronostic cognitif et comportemental

Dans les atteintes des réseaux cérébraux non primaires (lésions corticales, des faisceaux de substance blanche et/ou d'autres structures comme le cervelet, etc.), les relations entre anomalies précoces et devenir fonctionnel sont beaucoup plus complexes à établir car le développement des enfants à moyen et long terme dépend de multiples facteurs autres que lésionnels : facteurs génétiques, contextuels personnels et environnementaux tels que les expériences sensorielles et stimulations précoces, contexte psychoaffectif, nutrition, statut socio-économique et éducatif des parents, etc. L'IRM clinique à ce jour présente ainsi une faible valeur prédictive positive pour les troubles cognitifs, du langage, du comportement, et autres troubles fréquemment observés dans cette population : troubles des apprentissages (« dys »), trouble de déficit de l'attention (avec ou sans hyperactivité) ou encore troubles du spectre autistique.

## Conclusion

Les récents progrès de l'IRM permettent maintenant de proposer des marqueurs quantitatifs pour caractériser plus finement les lésions et anomalies cérébrales du nouveau-né dans les suites d'une agression pré- ou périnatale, et les utiliser pour affiner le pronostic fonctionnel. L'établissement précoce du pronostic moteur devient aujourd'hui plus fiable, mais prédire le devenir de l'enfant dans les autres domaines reste encore difficile avec les marqueurs actuellement disponibles. C'est pourquoi les recherches visent à développer de nouveaux marqueurs plus précis, par exemple en IRM à ultra-haut champ magnétique (7T) ou par le biais d'approches d'intelligence artificielle. Néanmoins, la prise en compte du contexte spécifique de chaque enfant reste cruciale pour une approche personnalisée, incluant son histoire clinique, ses caractéristiques environnementales et familiales, ses facteurs de risque mais aussi des facteurs protecteurs comme l'accès à un suivi médico-social personnalisé et à une rééducation précoce adaptée.

## Pour en savoir plus